\documentclass[pre,twocolumn,showpacs,preprintnumbers,amsmath,amssymb]{revtex4-1}

\pdfoutput=1

\usepackage{graphicx}
\usepackage{dcolumn}
\usepackage{bm}

\begin{document}

\newcommand{\bea}{\begin{eqnarray}}
\newcommand{\eea}{  \end{eqnarray}}
\newcommand{\bit}{\begin{itemize}}
\newcommand{\eit}{  \end{itemize}}

\newcommand{\be}{\begin{equation}}
\newcommand{\ee}{\end{equation}}
\newcommand{\ra}{\rangle}
\newcommand{\la}{\langle}
\newcommand{\U}{\widetilde{U}}


\def\bra#1{{\langle#1|}}
\def\ket#1{{|#1\rangle}}
\def\bracket#1#2{{\langle#1|#2\rangle}}
\def\inner#1#2{{\langle#1|#2\rangle}}
\def\expect#1{{\langle#1\rangle}}
\def\e{{\rm e}}
\def\proj{{\hat{\cal P}}}
\def\tr{{\rm Tr}}
\def\H{{\hat H}}
\def\Hdag{{\hat H}^\dagger}
\def\Lop{{\cal L}}
\def\Ehat{{\hat E}}
\def\Edag{{\hat E}^\dagger}
\def\Shat{\hat{S}}
\def\Sdag{{\hat S}^\dagger}
\def\Ahat{{\hat A}}
\def\Adag{{\hat A}^\dagger}
\def\U{{\hat U}}
\def\Udag{{\hat U}^\dagger}
\def\Zhat{{\hat Z}}
\def\Phat{{\hat P}}
\def\Op{{\hat O}}
\def\id{{\hat I}}
\def\x{{\hat x}}
\def\P{{\hat P}}
\def\Px{\proj_x}
\def\Pr{\proj_{R}}
\def\Pl{\proj_{L}}


\title{Classical properties of the leading eigenstates of quantum dissipative systems}

\author{Gabriel G. Carlo}
\affiliation{Departamento de F\'\i sica, CNEA, Libertador 8250,
(C1429BNP) Buenos Aires, Argentina}
\author{Leonardo Ermann}
\affiliation{Departamento de F\'\i sica, CNEA, Libertador 8250,
(C1429BNP) Buenos Aires, Argentina}
\author{Alejandro M. F. Rivas}
\affiliation{Departamento de F\'\i sica, CNEA, Libertador 8250,
(C1429BNP) Buenos Aires, Argentina}
\author{Mar\'\i a E. Spina}
\affiliation{Departamento de F\'\i sica, CNEA, Libertador 8250,
(C1429BNP) Buenos Aires, Argentina}

\email{carlo@tandar.cnea.gov.ar,ermann@tandar.cnea.gov.ar,rivas@tandar.cnea.gov.ar,spina@tandar.cnea.gov.ar}

\date{\today}

\pacs{05.45.Mt, 03.65.Yz, 05.45.a}

\begin{abstract}

By analyzing a paradigmatic example of the theory of dissipative systems -- the classical and quantum  dissipative
standard map -- we are able to explain the main features of the decay to the quantum equilibrium state.
The classical isoperiodic stable structures typically present in the parameter space of these kind of systems
play a fundamental role. In fact, we have found that the period of stable structures that are near in this space
determines the phase of the leading eigenstates of the corresponding quantum superoperator. Moreover,
the eigenvectors show a strong localization on the corresponding periodic orbits (limit cycles). We show that this 
sort of scarring phenomenon (an established property of Hamiltonian and projectively open systems) is present in 
the dissipative case and it is of extreme simplicity.

\end{abstract}

\maketitle

\section{Introduction}
\label{sec1}

Open quantum systems have received great attention recently
\cite{Weiss}. One of the main reasons has been the development of
quantum information and computation \cite{Nielsen,Preskill}, but
also the experiments with cold atoms \cite{CAexp,AOKR}, and
Bose-Einstein condensates \cite{BEC} can significantly profit from
their study. The engineering of reservoirs has been applied to
generate robust quantum states in the presence of decoherence
\cite{Kienzler}, providing with a new perspective in our study of
quantum environments. The route to chaos that is a typical feature
of dissipative systems has recently been studied in optomechanics
\cite{Bakemeier}. The interest has also been focused on many
body systems \cite{Hartmann}. In this case the study of the rocked
open Bose-Hubbard dimer has revealed the connection between the
interactions and bifurcations in the mean field dynamics. 
Very recently, quantum bifurcation diagrams have been 
obtained \cite{Ivanchenko}. In all
these established and promising areas the properties of the
leading eigenstates of the associated quantum superoperators are
of the utmost relevance, but have not been deeply investigated.

Up to now, the spectral behaviour of quantum dissipative systems
has been characterized by means of a Weyl law \cite{Spina} whose
theoretical explanation is still more elusive than the one for the
usual fractal Weyl law for projectively open systems. On the other
hand, asymptotic states associated to an eigenvalue $\lambda_{0}=1$
have been investigated in a system that is of interest in directed
transport studies, the dissipative modified kicked rotator map
\cite{Carlo,Beims}. However, the results obtained can be assumed
to be of general nature and thus applicable to any kind of
dissipative system. There, the fundamental role played by the so
called isoperiodic stable structures (ISSs) of the classical
system has been investigated. In particular it was conjectured
that their quantum counterparts (qISSs) have the simple shape of
the ISSs only for exceptionally large regular structures. In the
majority of the cases   qISSs look approximately the same as the
quantum chaotic attractors that are at their vicinity in parameter
space \cite{Carloprl}. This conjecture was proven in \cite{Carlo}
by means of a systematic exploration of the quantum parameter
space, showing how sharp classical borders become blurred and
neighboring areas influence each other through quantum
fluctuations. This phenomenon was called {\em parametrical
tunneling}.

We here investigate the other aspect of this phenomenon, that is,
how the presence of an ISS, even if not visible in the smooth
quantum parameter space, manifests itself in the quantum dynamics.
We focus on  the properties of the leading eigenstates. These
states , about which little is known, are very important since
they rule the transitory behaviour, i.e. the way the system decays
towards the equilibrium. It turns out that their phase space
structure as well as the phase of their corresponding eigenvalues
can be related to the shape and periodicity of a neighbouring ISS, 
respectively. The corresponding limit cycles not only influence 
the quantum system at the parameter values at which they appear classically, 
but also at their neighbourhood. This leads to localization on unstable 
periodic orbits. Hence, this phenomenon ubiquitous in the quantum 
chaos literature dealing with Hamiltonian systems, and also in the optics area 
that treats projectively open resonators \cite{scarring}, is also present 
in the dissipative arena, adopting a very simple behaviour.

For our calculations we choose a paradigmatic case of dissipative
dynamics: the classical and quantum dissipative standard map
(DSM). The corresponding parameter space (in the parameter range
we are considering) consists of a large regular region and a
chaotic sea where regular structures are embedded.

We explain the details of the classical and quantum
dissipative standard map in Sec. \ref{sec2}, together with the
techniques used to study their properties. In Sec. \ref{sec3} we
present our results focusing on the details of the parameter space
and taking special care of the Husimi representations of the
eigenvectors associated to the leading eigenstates. We close with
Sec. \ref{sec4} where we state our conclusions.

\section{The dissipative standard map: calculation methods}
\label{sec2}

The standard map can be thought of as describing the dynamics of a particle moving
in a coordinate $x$ with [$x\in(-\infty,+\infty)$] that is periodically kicked by
the single harmonic potential:
\begin{equation}
V(x,t)=k\left[\cos(x)\right]
\sum_{m=-\infty}^{+\infty}\delta(t-m \tau),
\end{equation}
where $k$ is the strength of these kicks and $\tau$ is the period.
Dissipation can be added to obtain the DSM \cite{Schmidt}
\begin{equation}
\left\{
\begin{array}{l}
\overline{n}=\gamma n +
k[\sin(x)]
\\
\overline{x}=x+ \tau \overline{n}.
\end{array}
\right.
\label{dissmap}
\end{equation}
We take $n$ as the momentum variable conjugated to $x$
and $\gamma$ ($0\le \gamma \le 1$) is the dissipation parameter.
By varying it the map performs a transition from the
Hamiltonian standard map ($\gamma=1$) to the one-dimensional circle map
($\gamma=0$). We can define a rescaled momentum variable $p=\tau n$
and the quantity $K=k \tau$ in order to simplify things.

This map can be easily quantized by taking: $x\to \hat{x}$, $n\to
\hat{n}=-i (d/dx)$ ($\hbar=1$). Since $[\hat{x},\hat{p}]=i \tau$
(where $\hat{p}=\tau \hat{n}$), the effective Planck constant is
$\hbar_{\rm eff}=\tau$. In order to reach the classical limit
$\hbar_{\rm eff}\to 0$, while $K=\hbar_{\rm eff} k$ remains
constant. We have taken $\hbar_{\rm eff}=0.042$, i.e. a finite
value. Quantum dissipation leads us to a master equation
\cite{Lindblad} for the density operator $\hat{\rho}$ in such a
way that
\begin{equation}
\dot{\hat{\rho}} = -i
[\hat{H}_s,\hat{\rho}] - \frac{1}{2} \sum_{\mu=1}^2
\{\hat{L}_{\mu}^{\dag} \hat{L}_{\mu},\hat{\rho}\}+
\sum_{\mu=1}^2 \hat{L}_{\mu} \hat{\rho} \hat{L}_{\mu}^{\dag} \equiv \Lambda \rho.
\label{lindblad}
\end{equation}
Here $\hat{H}_s=\hat{n}^2/2+V(\hat{x},t)$ is the system
Hamiltonian, \{\,,\,\} is the anticommutator, and $\hat{L}_{\mu}$ are the Lindblad operators
given by \cite{Dittrich, Graham}
\begin{equation}
\begin{array}{l}
\hat{L}_1 = g \sum_n \sqrt{n+1} \; |n \rangle \, \langle n+1|,\\
\hat{L}_2 = g \sum_n \sqrt{n+1} \; |-n \rangle \, \langle -n-1|,
\end{array}
\end{equation}
with $n=0,1,...$ and $g=\sqrt{-\ln \gamma}$ (to comply with the Ehrenfest theorem).

In order to perform the classical evolution we directly use the
map of Eq. \ref{dissmap}. In some cases, for comparison purposes,
we will be interested in a coarse grained version of the exact
dynamics \cite{UlamDMaps}. For this we will use the Ulam method
\cite{Ulam} which is an approximation to the Perron-Frobenius
operator arising from the Liouville equation for the the map in
Eq. \ref{dissmap}, obtained by discretizing the phase space.

In the quantum case we numerically integrate Eq. \ref{lindblad}
and obtain the evolution of the density matrix (symbolically) as
$\rho_{t+1} = e^ {\Lambda} \rho_{t}$, where $e^ {\Lambda}$ is a
non-unital superoperator of dimension $N^2 \times N^2$. The
effective Planck size is given by $\hbar_{\rm eff} \propto {1 /
N}$. Classical and quantum dissipation provides with a natural
bound that allows for the truncation of the phase space, leaving
all the relevant dynamics inside of the resulting domain, and
providing with finite (super)operators. The diagonalization of the
Ulam superoperator and of the quantum $e^{\Lambda}$, is performed
by the Arnoldi method \cite{Arnoldi}.

\section{Properties of the leading eigenstates: the decay towards equilibrium}
\label{sec3}

\begin{figure}[htp]
 \includegraphics[width=0.47\textwidth]{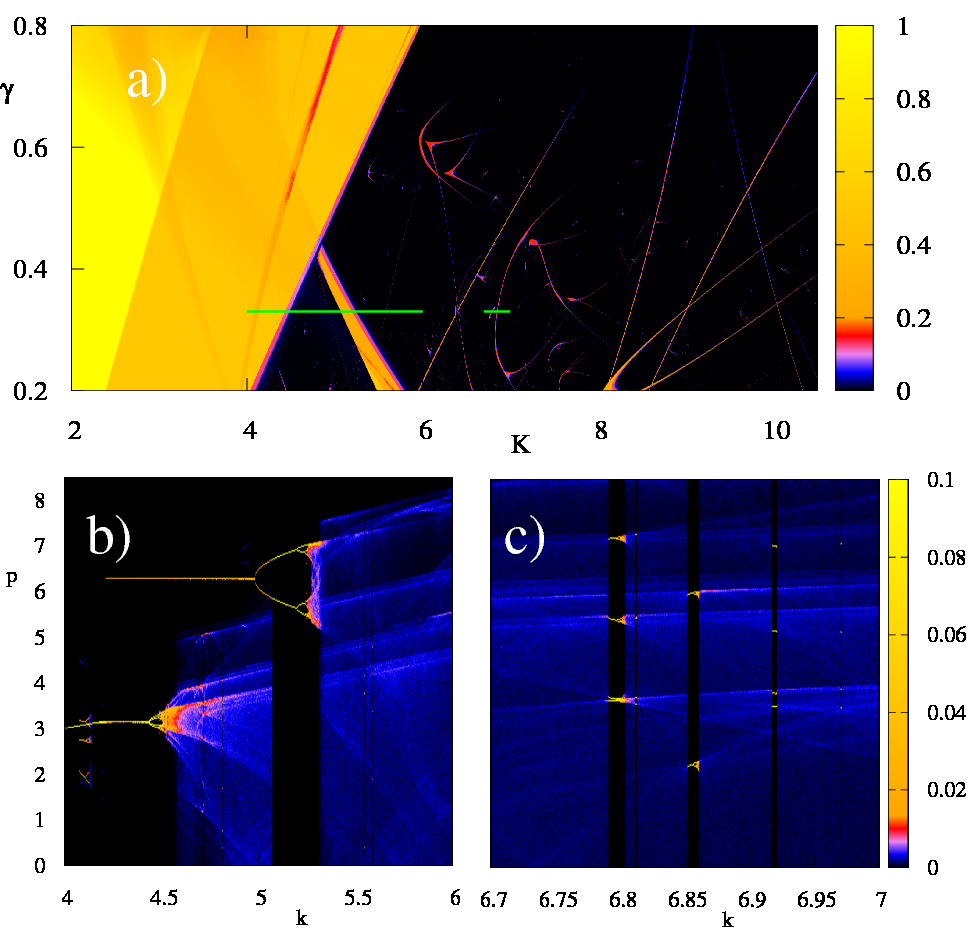}
\caption{(color online) Top panel $a)$ shows the 
participation ratio $\eta$ vs. parameters $k$ and $\gamma$ (for details see main text). 
Bottom panels $b)$ and $c)$ illustrate
probability in momentum $P(p)$ as a function of $k$ for
$\gamma=0.33$ corresponding to the ranges shown in green lines in
panel $a)$. }
 \label{fig1}
\end{figure}

In order to characterize the behaviour of the leading eigenstates
of the DSM we need first to explore the parameter space spanned by
$k$ and $\gamma$. For that purpose we calculate the participation
ratio defined by $\eta=(\sum_iP(p_i)^2)^{-1}/N$. $P(p_i)$ is a
discretized limiting momentum ($p$) distribution, taken after
evolving $10000$ time steps a bunch of $10000$ random initial
conditions in the $p=[-k/(1-\gamma);k/(1-\gamma)]$ band of the cylindrical phase
space (i.e., the trapping region defined in \cite{Martins}). 
We have taken a number of bins given by a Hilbert space 
dimension $N=1000$. Taking a finer coarse-graining would not
affect the main properties in which we are interested since the
distance among points of the simple limit cycles is always greater
than the bin size. This measure is a good indicator of the
complexity of the asymptotic distribution. It is worth mentioning 
that the DSM contains multistability regions with a great number 
of coexisting attractors \cite{Martins}. However, we are interested 
in clearly identifying those regions where just regular behavior is found 
and those where a chaotic attractor is present. For this objective 
our measure is a very suitable one. Results are shown in
Fig. \ref{fig1} a).

The diagonal line that can be clearly seen in this figure
corresponds to the onset of chaotic behavior. On the right side of
this line, in general tiny ``regular islands'' (where only regular behaviour is present) 
can be found within the ``chaotic
sea''. These are the previously mentioned ISSs, i.e. Lyapunov
stable islands that come organized into families in the parameter
space, which usually have long antenna like branches leading to
nicknaming them as ``shrimps''.

The bifurcations diagrams along a line with fixed dissipation
parameter $ \gamma = 0.33 $ and $k$ in the intervals $ 4.0 \le k
\le 6.0 $ (Region 1) and $ 6.7 \le k \le 7.0 $ (Region 2) are
shown in panels b) and c) of Fig. \ref{fig1} (we display the 
probability in momentum $P(p)$ as a function of $k$). They complement Fig.
\ref{fig1} a) by providing information on the limit cycles
which characterize the selected regular regions.

\begin{figure}[htp]
 \includegraphics[width=0.47\textwidth]{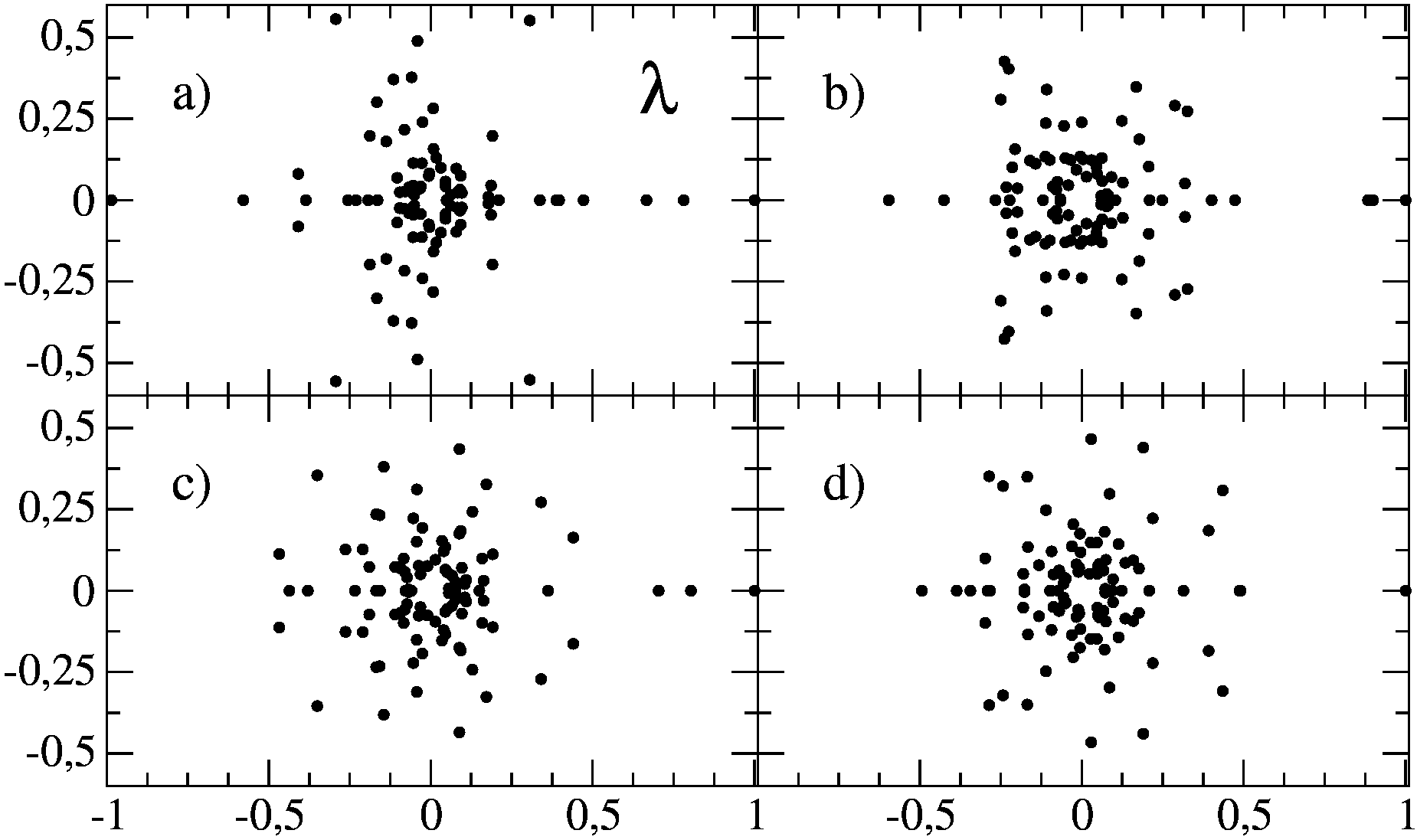}
\caption{100 largest eigenvalues of the quantum superoperator
$e^{\Lambda}$ for a) ($ \gamma = 0.33,k=4.0$), b) ($ \gamma =
0.33,k=5.13 $), c) ($ \gamma = 0.33,k=5.5 $), d) ($ \gamma =
0.33,k=6.0 $). $\hbar_{\rm eff}= 0.042$.}
 \label{fig2}
\end{figure}

\begin{figure}[htp]
 \includegraphics[width=0.47\textwidth]{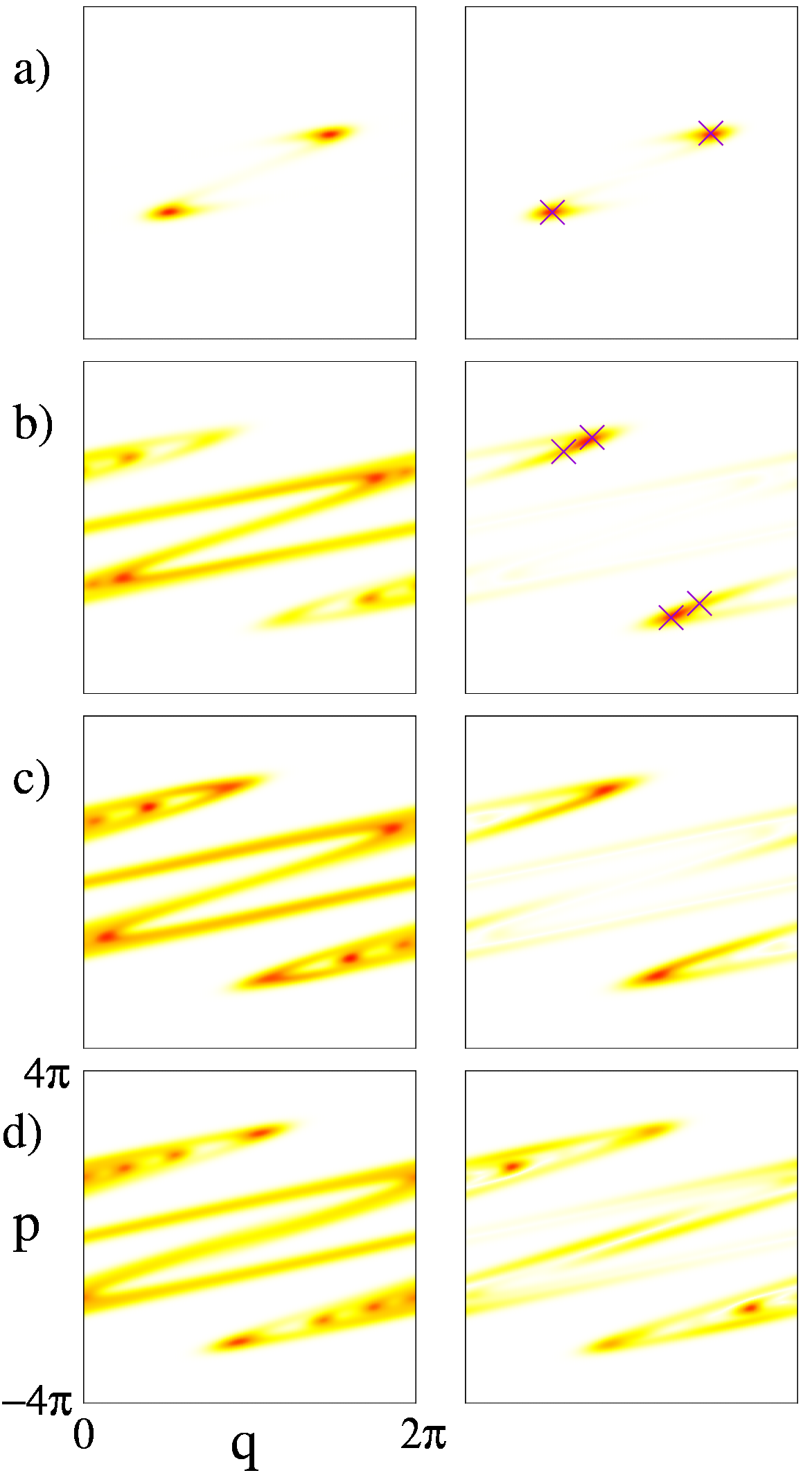}
\caption{(color online) Husimi representation of the invariant
state (left panel) and leading eigenstate (right panel) for  a) ($
\gamma = 0.33,k=4.0$), b) ($ \gamma = 0.33,k=5.13 $), c) ($ \gamma
= 0.33,k=5.5 $), d) ($ \gamma = 0.33,k=6.0 $). $\hbar_{\rm eff}= 0.042$.
In the right panels of a) and b) the periodic points of the ISSs
are marked with crosses.}
 \label{fig3}
\end{figure}

We first present the results of the diagonalization of the quantum
superoperator $e^{\Lambda}$ with an effective $\hbar_{\rm eff}= 0.042$
for four representative points of Region 1. Case a) corresponds to
($ \gamma = 0.33,k=4.0$) located in the regular region, case b) to
($ \gamma = 0.33,k=5.13 $) in the interior of the largest regular
island found in the explored parameter space. The two other points
belong to the chaotic set: ($ \gamma = 0.33,k=5.5 $) (case c) ) is
in the vicinity of the island, while ($ \gamma = 0.33,k=6.0 $)
(case d) ) lies in a region where no islands are visible at our
resolution. Fig. \ref{fig2} shows the eigenvalue spectra in
complex space for these four cases. The corresponding
Husimi representations of the invariant state (left panels) and of
the leading eigenstate (right panels) are displayed in Fig.
\ref{fig3}.

The invariant state shown in \ref{fig3} a) for parameters deep
inside the regular region exhibits a point like structure which
coincides, within quantum uncertainty, with a stable periodic
orbit with winding number $ 0/2 $. This orbit has been calculated
with the method proposed in \cite{wenzel} and is marked with
crosses. It corresponds to the period two limit cycle which is
visible in the bifurcation diagram of Fig. \ref{fig1} b). The
leading resonance with $ \lambda_1= -0.985733 $ shown in the right
panel of Fig. \ref{fig3} a) presents a very similar point-like
distribution. Moreover we have verified that all states
corresponding to eigenvalues close to the unit circle in the
spectrum of Fig. \ref{fig2} a) have a simple structure. For real
eigenvalues  the Husimi distributions look very much the same as
the one of Fig. \ref{fig3} a) while for the largest complex
eigenvalues which have phases approximately equal to multiples of
$ \pi / 3 $ they are peaked on a period three limit cycle also
present in this region (see Fig.\ref{fig1}  b)). As we go to lower
eigenvalues of the spectrum all point-like eigenstates get more
and more blurred. These results confirm that in the regular region
of the parameter space quantum dynamics reproduces (within quantum
uncertainty) the main features of the classical one.

This is not the case for parameters belonging to an ISS. As shown
in Fig.\ref{fig3}  b) the invariant state does not follow the
asymptotic classical distribution characterized by a period two
limit cycle, but has instead the complex structure of a strange
attractor. In fact, it looks very much the same as the states
obtained in the chaotic region which are shown in the two panels
below. This is an indication  that in what regards the equilibrium
states the ISS cannot be resolved at our value of $\hbar_{\rm eff}$.
However the situation is different if we look at the leading
eigenstate shown in the right panel of Fig.\ref{fig3}  b). Its
Husimi distribution is strongly localized on a periodic orbit with
winding number $ 2/2 $ marked by crosses in this Figure. The
corresponding eigenvalue is real and large ( $ \lambda_1 \sim 0.9
$ ) as we can see from the spectrum of Fig.\ref{fig2}  b), hinting
at a long decay time towards the complex invariant state. The
following states are significantly shorter lived ($ |\lambda_1|
\sim 0.6$) and even if they show some localization on orbits present
in the island, they essentially have the structure of the
attractor.

We obtain analogous results for parameters which do not belong to
the ISS but are close to it. As expected, the equilibrium state
shown in Fig.\ref{fig3}  c) is a complex attractor, this time in
agreement with the asymptotic classical distribution. However, as
in the previous case, the presence of the regular island
influences the leading eigenstate. Even if the distribution shown
in the right panel of Fig.\ref{fig3}  c) is not strictly
point-like it still has a clearly enhanced probability on the
periodic orbit $ 2/2 $ which is unstable now. In fact, the enhancement of probability 
on an unstable periodic orbit is the definition of scarring.  The
corresponding eigenvalue $ \lambda_1= 0.8 $ is real and fairly
large, indicating a long transient also in this case. The spectrum
in  Fig.\ref{fig2}  c) shows the existence of a second real
eigenvalue $ \sim 0.7 $ considerably larger than the radius of the
dense disk. It corresponds to a state with a similar scarred
structure.

Finally, the calculation for parameters further away from regular
islands gives the typical features of chaotic dynamics. Both the
equilibrium state and the leading resonance in Fig.\ref{fig3}  d)
are quantum chaotic attractors.  No exceptional long-lived states
exist. The spectrum of Fig.\ref{fig2}  d) shows the presence of a
considerable gap between eigenvalue 1 and the dense disk of
eigenvalues ($ |\lambda_1|= 0.5 $).

\begin{figure}[htp]
 \includegraphics[width=0.47\textwidth]{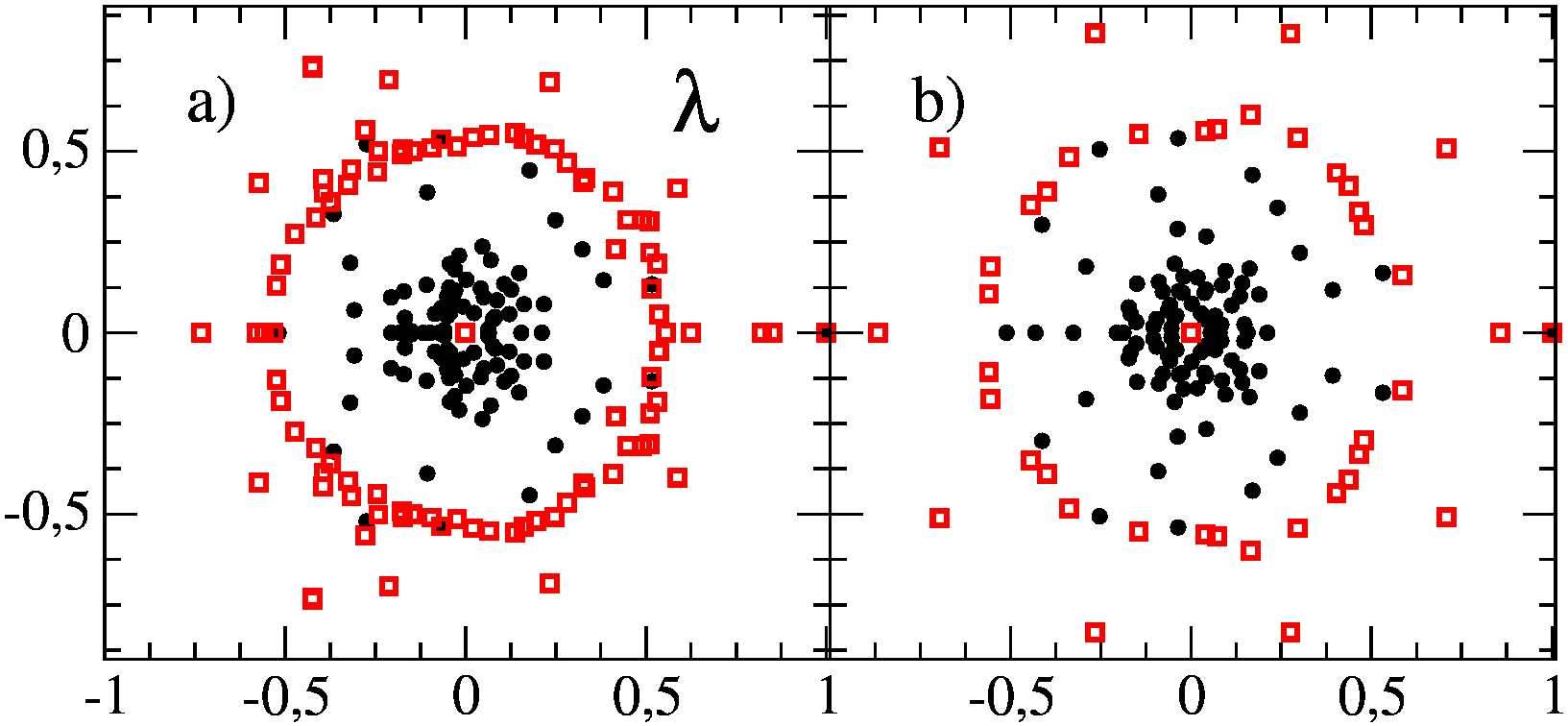}
\caption{(color online) Largest eigenvalues of the quantum
superoperator $e^{\Lambda}$ (with $\hbar_{\rm eff}= 0.042$) and of the
Perron-Frobenius superoperator (with $N=300$) for a) ($ \gamma =
0.33, k=6.86 $), b) ($ \gamma = 0.33,k=6.8 $). Black dots
correspond to the quantum model while (red) gray squares to the
classical one. }
 \label{fig4}
\end{figure}

\begin{figure}[htp]
 \includegraphics[width=0.47\textwidth]{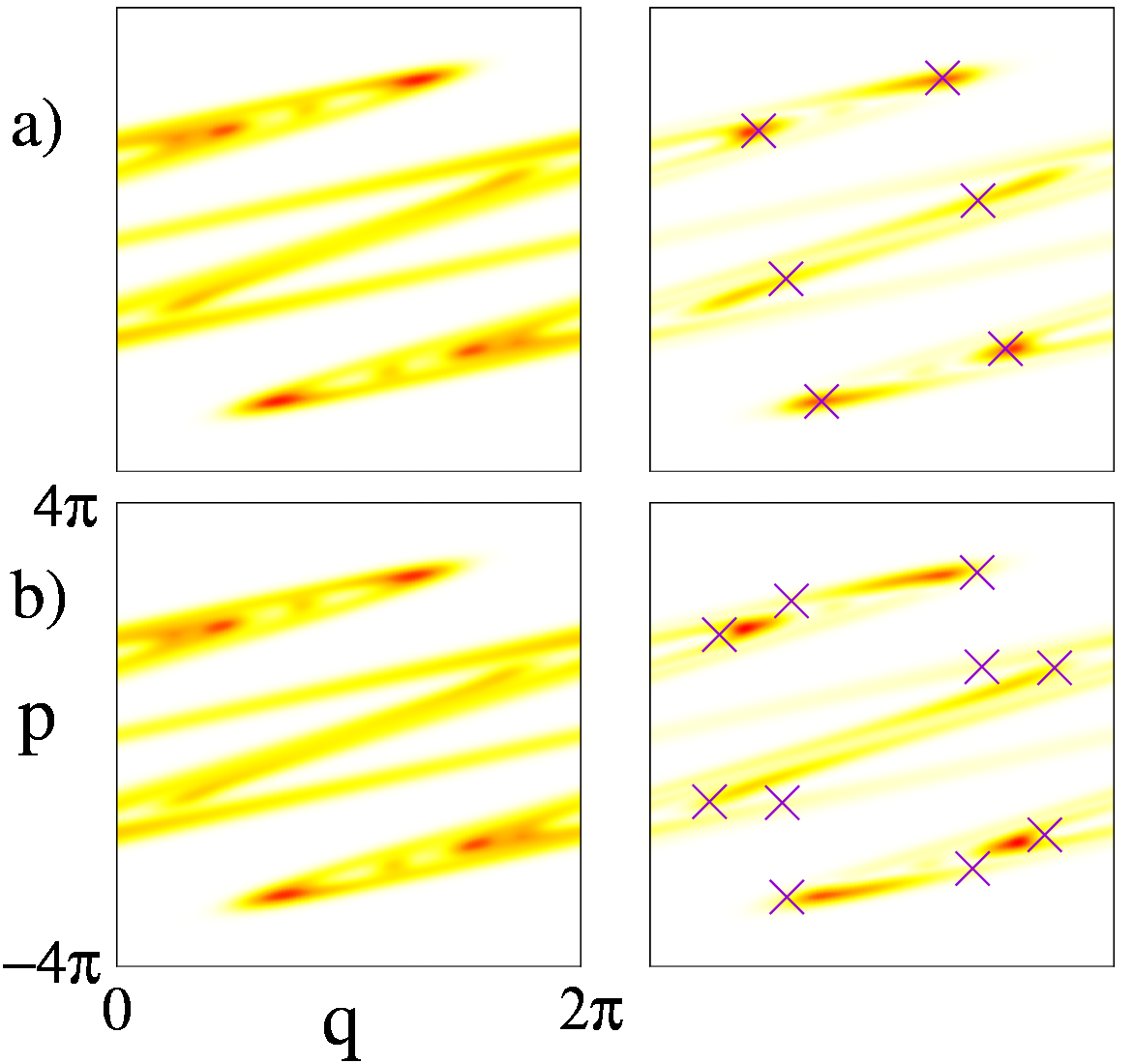}
\caption{(color online) Husimi representation of the invariant
state (left panel) and leading eigenstate (right panel) for a) ($
\gamma = 0.33, k=6.86 $), b) ($ \gamma = 0.33,k=6.8 $). $\hbar_{\rm eff}=
0.042$. In the right panels the periodic points of the ISSs are
marked with crosses.}
 \label{fig5}
\end{figure}

We now analyze some examples in Region 2 where only small ISSs
exist, hardly visible in parameter space. We focus on the largest
ones recognizable in the bifurcation diagram of Fig.\ref{fig1} c),
one centered at $ k=6.86 $ and the other at $ k=6.8 $. They are
characterized by a period three and a period ten limit cycles
respectively. The results of the diagonalization of the quantum superoperator
for  a) ($ \gamma = 0.33, k=6.86 $) and b) ($ \gamma = 0.33,k=6.8
$) are shown in Figs. \ref{fig4} and \ref{fig5}. For comparison
purposes the spectra of the Perron-Frobenius superoperator
corresponding to the classical map for these parameters are also
indicated in Fig. \ref{fig4} .

The Husimi distribution of the invariants displayed in the left
panels of Fig. \ref{fig5} look very similar in both cases and have
the structure of strange attractors, thus differing from the
asymptotic point-like classical distributions. This is to be
expected since, as we have seen in the previous cases, the quantum
equilibrium distributions are not affected by the presence of
regular regions unless these are really large. For their part, the
spectra shown in Fig. \ref{fig4} also present at first sight the
characteristic features of chaotic behavior, in the sense that
they both have a well defined gap (associated with the decay time
towards the invariant) between $ \lambda = 1 $ and the disk where
the remaining eigenvalues concentrate. However if we  look closer
at the spectrum of Fig. \ref{fig4}a) we observe that the complex
phases of the leading eigenvalues (with $ |\lambda | \sim 0.59 $)
are multiples of $ \sim {2 \pi / 3} $, just as the ones in the
corresponding Perron-Frobenius spectrum. This quantum classical
correspondence is confirmed if we look at the structure of the
corresponding eigenstate shown in the right panel of Fig.
\ref{fig5} a) which is strongly scarred by the periodic orbit with
winding number $ 2 / 3 $ (indicated with crosses)
characterizing the island.

We now focus on the spectrum of Fig. \ref{fig4} b) . In this case
the limit cycle has period ten, and this periodicity manifests
itself in the longest-lived eigenvalues of the Perron-Frobenius
spectrum, which have phases equal to the tenth roots of unity.
However no traces of this periodic 10-family are present in the
quantum spectrum nor in the structure of the leading eigenstate of
Fig. \ref{fig5} b). In fact, the distribution of this state is
very similar to the one shown in the panel above, i.e., it is
peaked on the period three orbit. Moreover it corresponds to an
eigenvalue with a phase $ \sim {2 \pi / 3} $ as in the
previous case. This seems to indicate that limit cycles with high periodicity
that clearly determine the coarse grained classical dynamics for a
given point in parameter space even at relatively low resolution
(in our case $N=300$) are not robust enough to manifest in the
quantum case. Rather, the quantum behavior is influenced by
regularity regions which might not be the closest ones in parameter
space, but correspond to shorter limit cycles.

\section{Conclusions}
\label{sec4}

The standard map is a paradigmatic model in the classical and the
quantum chaos literature, while the dissipative version shares the
same status for the theory of dissipative systems. 
Its classical parameter space shows a rich structure, proper of
nonintegrable dissipative systems, which consists of a large regular
region and a chaotic domain in which ISSs are
embedded. As it has been established in previous works, these
small stable structures are washed away by quantum fluctuations,
implying that the equilibrium states (in regions other the regular
region) have the structure of chaotic attractors, independently of
their location in the chaotic sea.

In this work we have investigated how the presence of an ISS
affects its neighborhood in parameter space by focusing on the
spectral properties of the quantum mechanical superoperator. 
We have found that the ISSs  play a fundamental role in the
morphology of the leading eigenvectors of spectra obtained for
parameters where they are present (at the classical level), but 
more interestingly, also at their vicinity. These eigenvectors turn out to be
particularly long-lived and show a clear scarring by a limit
cycle that belongs to the ISS. The second effect concerns the
complex phases of the corresponding eigenvalues which are related
to the periodicity of these limit cycles. 

Our results strongly suggest that three
factors are involved in the appearance of this mechanism (which shows another aspect of the 
parametrical tunneling). In the first place, 
the distance of the considered parameter space location to the ISS should be short 
in terms of the distance to other structures. Secondly, its size in parameter space 
should be greater or the same than other neighbouring regions. Finally, the period 
of the related periodic orbit should be short enough to be perceived by quantum 
mechanics (in fact, this is also an important 
property in order to even find them in the classical explorations \cite{Martins}).
We conjecture that, given the paradigmatic nature of our model, these results can be
considered to be generic properties of quantum dissipative
systems.

\section*{Acknowledgments}

Support from CONICET under project PIP 112 201101 00703 is gratefully acknowledged. One of us (LE) acknowledges 
support from ANPCYT under project PICT 2243-(2014).

\vspace{3pc}


\end{document}